\documentstyle[12pt]{article}

\begin{document}
\setlength{\parindent}{0pt}
\begin{center}
\begin{large}
{\bf Tensor forces and relativistic corrections in quarkonium}\\
\end{large}
\vspace*{1cm}
{\bf I. I. Haysak $^{\ast}$} and {\bf V. S. Morokhovych $^{\dag}$}  \\
Department of Theoretical Physics, Uzhgorod National University,
Voloshyna Street 32, 88000 Uzhgorod,
Ukraine \\

$ \begin{array}{llll} ^{\ast}\mbox{{\it e-mail:}} &
 \mbox{haysak@univ.uzhgorod.ua} & ^{\dag}\mbox{{\it e-mail:}} &
 \mbox{morv@univ.uzhgorod.ua}
\end{array}
$

{\bf S. Chalupka} \\
Department of Theoretical Physics and Geophysics, Faculty of
Science, P. J. {\v S}af{\'a}rik University,  Moyzesova 16, 041  54
Ko{\v s}ice, Slovak Republic \\

$ \begin{array}{ll} \mbox{{\it e-mail:}} &
 \mbox{chalupka@kosice.upjs.sk}
\end{array}
$

{\bf M. Sal\'ak} \\
Department of Physics Pre{\v s}ov University, 17 novembra 1,
08009 Pre{\v s}ov, Slovak Republic \\
\end{center}
\vspace*{2cm} {\bf  Abstract.} The influence of tensor forces on
hyperfine splitting and decay widths of mesons are investigated
within the limits of quasirelativistic potential quark model. As
the starting the system of the Rarita-Schwinger equations is
used. We use a potential motivated by QCD with a mixed Lorentz structure.
\vspace*{0.5cm} \\
{\bf 1. Introduction}
\vspace*{0.5cm} \\
\hspace*{0.5cm} As well known, the heavy meson spectrum can be
successfully obtained starting from a potential model [1-3]. The
describing of the fine and hyperfine structure including decay
widths is problematic in framework of unified approach. Many
authors with different types potentials for quark-antiquark
interaction describe only separate properties of mesons.
Following papers [4-6] we analyse the Lorentz structure of the
potential with reference to $u \bar u$ system, charmonium and
bottonium hyperfine splitting. It is well known that tensor
forces bring about mix wave with different angular momentum [7].
Majority of authors consider that contribution of tensor forces
in spectrum of bound states of quarkonia is negligible [8,9]. But,
it was shown [3] that $D$ waves give 1 - 5 \% contribution for
energy spectrum in ground triplet
state. \\
\hspace*{0.5cm} In the present paper we will use model which is
based on the quasirelativistic Breit-Fermi Hamiltonian [10] and
we will be analyze an influence of tensor forces with reference
to two-quark system. Singlet states  are described by
Schr\"odinger equation and for triplet states we use the system
of Rarita-Schwinger equations [7].
\vspace*{0.5cm} \\
{\bf 2. Hyperfine splitting}
\vspace*{0.5cm} \\
\hspace*{0.5cm} It is widely accepted that the interaction
between two quarks (or heavy quark - antiquark) consist of a
short range part describing the one-gluon-exchange and a
infinitely rising long-range part responsible for confinement of
the quarks. This also gives good approximation to the lattice
potential. We use
\begin{equation}
V_0 = V_{V} (r) + V_{S} (r), \label{eq1}
\end{equation}
where
\begin{equation}
V_{V} = - \frac{\alpha(r)}{r}, \quad   V_S (r) = kr. \label{eq2}
\end{equation}
(The color factor 4/3 has been absorbed in the definition of
$\alpha$). Wilson loop techniques suggest that the confining
potential should be taken purely scalar, but relativistic
potential calculations which have been published [11-13] show a
need for (some) vector confinement. Maintaining $V_{V} (r) +
V_{S}(r)$ unchanged we do allow a fraction of vector confinement
[4,6,13]. Thus
\begin{equation}
V_{V}(r) = - \frac{\alpha(r)}{r} + \beta_{V} r, \quad V_S (r) =
\beta _S r \quad (\beta_{V} + \beta_S = k). \label{eq3}
\end{equation}
Then
\begin{equation}
V_0 = (- \frac{\alpha(r)}{r} + \beta_{V} r) + \beta_S r.
\label{eq4}
\end{equation}
The confining potential transforms as the Lorentz scalar and
vector potential transforms as the time component of a
four-vector potential. As we can see, the choice of Lorentz
structure of potential for quark-antquark interaction is
important model for study of spin effects [13-17]. \\
\hspace*{0.5cm} We consider the Breit-Fermi Hamiltonian (in case
of equal masses $m_{1}=m_{2}=m$ ):
\begin{equation}
H =\frac{\overrightarrow{p}^{2}}{m} + V_{0} + H_{LS} + H_{SS} +
H_{T} \label{eq5}
\end{equation}
with spin-orbit term
\begin{equation}
H_{LS} = \frac{1}{2m^{2} r}[ 3 \frac{dV_{V}}{dr} -
\frac{dV_S}{dr}] (\mbox{\boldmath $L$}.\mbox{\boldmath $S$}),
\label{eq6}
\end{equation}
$\mbox{\boldmath $S$} = \mbox{\boldmath $S$}_1 + \mbox{\boldmath
$S$}_2$ is the total spin of bound state and $\mbox{\boldmath
$L$} = \mbox{\boldmath $r$} \times \mbox{\boldmath $p$}$ is the
relative orbital angular momentum of its constituents, the
spin-spin term
\begin{equation}
H_{SS} = \frac{2}{3 m^2} \Delta V_V (r) \mbox{\boldmath $S$}_1 .
\mbox{\boldmath $S$}_2, \label{eq7}
\end{equation}
and the tensor term
\begin{equation}
H_T = \frac{1}{12 m^2} [\frac{1}{r} \frac{d V_V}{dr} - \frac{d^2
V_V}{dr^2}] S_{12}, \label{eq8}
\end{equation}
where
\begin{equation}
S_{12} = 12[(\mbox{\boldmath $S$}_1 . \mbox{\boldmath $n$})
(\mbox{\boldmath $S$}_2 . \mbox{\boldmath $n$}) - (
\mbox{\boldmath $S$}_1 . \mbox{\boldmath $S$}_2) / 3] =
3(\mbox{\boldmath $\sigma$}_1 . \mbox{\boldmath $n$})
(\mbox{\boldmath $\sigma$}_2 . \mbox{\boldmath $n$}) -
(\mbox{\boldmath $\sigma$}_1 . \mbox{\boldmath $\sigma$}_2),
\label{eq9}
\end{equation}
$$\mbox{\boldmath $n$} = \mbox{\boldmath $r$} / r, \quad
\mbox{\boldmath $S$}_i = \mbox{\boldmath $\sigma$}_i / 2.$$ For
bound-state constituents of spin $S_1 = S_2 = 1/2$, the scalar
product of their spin, $\mbox{\boldmath $S$}_1 . \mbox{\boldmath
$S$}_2$, is given by
\[ \mbox{\boldmath $S$}_1 . \mbox{\boldmath $S$}_2 =
\left\{ \begin{array} {ll}
-\frac{3}{4} & \mbox{for spin singlets}, S = 0, \\
+\frac{1}{4} & \mbox{for spin triplets}, S = 1.
\end{array}
\right. \] Taking picture (\ref{eq4}), then (\ref{eq6}),
(\ref{eq7}) and (\ref{eq8}) yield
\begin{equation}
H_{LS} = \frac{1}{2 m^2} [\frac{3 \alpha}{r^3} + \frac{3 \beta_V
- \beta_S}{r}] (\mbox{\boldmath $L$} . \mbox{\boldmath $S$}),
\label{eq10}
\end{equation}
\begin{equation}
H_{SS} = \frac{2}{3 m^2} [4 \pi \alpha \delta (\mbox{\boldmath
$r$}) + \frac{2 \beta_V}{r}] (\mbox{\boldmath $S$}_1 .
\mbox{\boldmath $S$}_2), \label{eq11}
\end{equation}
\begin{equation}
H_T = \frac{1}{12 m^2} (\frac{3 \alpha}{r^3} + \frac{\beta_V}{r})
S_{12}. \label{eq12}
\end{equation}
Two-quark system may exist in the spin singlet and the triplet
states (Table 1).
\begin{table}[1]
\caption{\bf Conditions of two fermion system}
\begin{center}
\begin{tabular}{|c|c|c|c|c|}
\hline & \multicolumn{2}{c|}{Singlet condition} &
\multicolumn{2}{c|}{Triplet condition} \\
& \multicolumn{2}{c|}{$(S=0)$} &
\multicolumn{2}{c|}{$(S=1)$} \\
\hline
$J/P$ & + & - & +  & - \\
\hline
0   &  - - - -  & $^1S_0$  & $ ^3P_0$  & - - - -  \\
\hline
1 & $^1P_1$ & - - - -     & $^3P_1$ & $^3S_1 + ^3D_1$ \\
\hline
2 & - - - - & $^1D_2$   & $^3P_2 + ^3F_2$ & $^3D_2$ \\
\hline
\end{tabular}
\end{center}
\end{table}

If $S = 0$, then necessarily $L = J$ and eigenfunction is of
the form $(f(r)/r) {\cal Y}^M_{J0J}$,\\
where
\begin{equation}
{\cal Y}^M_{JLS} = \sum_{m \mu} \langle LS m \mu | JM \rangle
\langle Y_{Lm} (\Omega)| S \mu \rangle. \label{eq13}
\end{equation}
Since
$$\mbox{\boldmath $S$} | 00 > = 0, $$
\begin{equation}
S_{12} = 2[3 (\mbox{\boldmath $S$} . \mbox{\boldmath $n$})^2 -
\mbox{\boldmath $S$}^2] \label{eq14}
\end{equation}
we get
$$S_{12} {\cal Y}^M_{JL0} \equiv S_{12} Y_{Jm}(\Omega)|00 >
= 0.$$ Consequently, $f(r)$ satisfies radial equation
\begin{equation}
\frac{d^2 f}{dr^2} + [k^2 - \frac{L(L + 1)}{r^2}- ^1U_c] f = 0,
\label{eq15}
\end{equation}
where $^1U_c = m ^1V_c ;   ^1V_c = V_V(r) + V_S (r) - (3/4)
V_{SS}$ and $k^2 = m E$. If $S = 1$ and $P = (-1)^{J+1}$, the
only possible values of $L$ are $J + 1$ (unless $J = 0$ in which
case
there is just the one value $L = 1$). \\
\hspace*{0.5cm} The wave function for ground triplet state of $q
\bar q$ system with negative parity $(P = -1)$ is a mixture of
state $^3S_1$ and $^3D_1$ and may be put in the form
\begin{equation}
\psi = \psi_S + \psi_D \equiv \frac{1}{r} u(r) {\cal Y}^1_{101} +
\frac{1}{r} w (r) {\cal Y}^1_{121}. \label{eq16}
\end{equation}
The input $D$-waves to the wave function of vector meson makes
part of percent. But the input $D$-wave to the spectrum of energy
makes some percent. It is explained by including of the
interference member to the energy spectra. Knowing the wave
function it is possible to calculate input of every  component to
the energy level. We have
$$ E = \langle \psi_S + \psi_D | H | \psi_S + \psi_D
\rangle.$$ Then the energy value can be split into components
$$E = E_S + E_D + E_{SD},$$
where $E_S, E_D$ and represent inputs of $S,D$-wave and of
the interference wave, respectively. \\
\hspace*{0.5cm} Now, for $S = 1$ state we have the wave function
(\ref{eq16}) we know that,
$$S_{12} {\cal Y}^1_{010} = \sqrt{8}{\cal Y}^1_{121}$$
and
$$S_{12} {\cal Y}^1_{121} = \sqrt{8}{\cal Y}^1_{101} -
2 {\cal Y}^1_{121}.$$ Futher
$$\mbox{\boldmath $L$}^2 {\cal Y}^1_{101} = 0, \quad
\mbox{\boldmath $L$}^2 {\cal Y}^1_{121} = 6 {\cal Y}^1_{121}.$$
Since
$$H = - \frac{1}{m} \frac{d^2}{dr^2} +
\frac{\mbox{\boldmath $L$}^2}{m r^2} + ^3V_c + V_{LS}
\mbox{\boldmath $L$} . \mbox{\boldmath $S$} + V_T S_{12},$$ and
$\mbox{\boldmath $S$}_1 . \mbox{\boldmath $S$}_2 = 1/4,
\mbox{\boldmath $L$} . \mbox{\boldmath $S$} = -3, S_{12} = - 2$
for $S = 1$. Then the equation $(H - E) \psi = 0$ is equivalent
to the Rarita-Schwinger system
$$[ - \frac{1}{m} \frac{d^2}{dr^2} - E + ^3V_c] u + \sqrt 8
V_T w = 0,$$
\begin{equation}
[ - \frac{1}{m} \frac{d^2}{dr^2} - E + \frac{6}{m r^{2}} + ^3 V_c
- 2 V_T - 3 V_{LS}] w + \sqrt 8 V_T u = 0, \label{eq17}
\end{equation}
where
$$^3V_c = V_V + V_S + \frac{1}{4} V_{SS}.$$
The system (\ref{eq17}) we rewrite in the matrix form
$$\widehat{h}{u \choose w} = E {u \choose w},$$
with
$$\widehat{h} = \widehat{h}_0 + \widehat{W}$$
and
\begin{equation}
\widehat{h}_{0} = \left ( \begin{array}{cc} \! - \displaystyle \frac{1}{m}
\frac{d^2}{\displaystyle dr^2} \! - \! \frac{\alpha}{r}\! + \! kr \! + \!
\frac{1}{m^2 r} \frac{\beta_V}{3}
& \displaystyle \frac{1}{m^2 r} \frac{\sqrt 2 \beta_V}{6} \\
\displaystyle \frac{1}{m^2 r} \frac{\sqrt 2 \beta_V}{6} & \!- \displaystyle \frac{1}{m}
\frac{d^2}{dr^2} \! + \frac{6}{m r^{2}}\!- \! \frac{\alpha}{r} \!
+ \! kr \! + \! \frac{1}{m^2 r} \!( - \! \frac{13}{3} \beta_V \! +
\! \frac{3}{2} \beta_S)
\end{array} \right ), \label{eq18}
\end{equation}
\begin{equation}
{\widehat{W}} = \left ( \begin{array}{cc}
\frac{1}{4} \Delta V_{SS} & \frac{\sqrt 8}{12} \Delta V_T \\
\frac{\sqrt 8}{12} \Delta V_T & \frac{1}{4} \Delta V_SS -
\frac{2}{12} \Delta V_T - 3 \Delta V_{LS}
\end{array} \right ), \label{eq19}
\end{equation}
where
$$\Delta V_{SS} = \frac{8 \pi \alpha}{3m^2}\delta(\vec r),
\quad \Delta V_{LS} = \frac{3 \alpha}{2 m^2 r^3}, \quad \Delta V_T
= \frac{3 \alpha}{ m^2 r^3}.$$ Then matrix elements of correction
terms are
\begin{eqnarray}
\nonumber \Delta E_{mn} &=& \int u_m (\frac{2 \pi \alpha}{3 m^2}
\delta(\mbox{\boldmath $r$})) u_n d r + \int w_m
(\frac{\alpha \sqrt 2}{2 m^2 r^3}) u_n d r  \\
&+& \int u_m (\frac{\alpha \sqrt 2}{2 m^2 r^3}) w_n d r + \int w_m
(\frac{2 \pi \alpha}{3 m^2} \delta(\mbox{\boldmath $r$}) -
\frac{5 \alpha}{m^2 r^3}) w_n d r. \label{eq20}
\end{eqnarray}
We put $\alpha = \frac{4}{3} \alpha_s$. The $QCD$ running
constant $\alpha_s$ is determined according to the formula

$$\alpha_s (q^2) = 12 \pi/ [(33 - 2 n_j) ln (q^2/\land^2)],$$

where $\land = \land_{QCD} = 140 MeV$; $n_j = 3$ for light and
mixed mesons; $n_j = 4$ for $c \bar c$- and $b \bar b$-
quarkonium. For Cornell potential hyperfine splitting of mesons
may be described with a rather accuracy
if $q = 2 \mu$ ($\mu$ - reduced mass) is chosen [18]. \\
\hspace*{0.5cm} In fact $\Delta E_{mn}$ have singularities
$(\delta (\mbox{\boldmath $r$}))$ and $r^{-3}$. Therefore one has
to introduce smeared  $\delta$ function and cut of for $r^{-3}$
to weaken singularities. Our calculation were made with tensor
forces and without tensor forces. We have used the following
parameters of potentials: $\beta_V = 0.001 GeV^2, \beta_S = 0.179
GeV^2$ for $u \bar u$-systems; $\beta_V = 0.04 GeV^2, \beta_S =
0.14 GeV^2$ for charmonium and bottonium, $\alpha_s (c \bar c) =
0.38, \alpha_s (b \bar b) = 0.24, \alpha_s (u \bar u) = 0.54$.
The quark  masses  is: $m_c = 1.4 GeV, m_b = 4.7 GeV$ and $m_u =
0.33 GeV$. In the works [13,16] it was pointed out that the best
agreement is obtained when  $\beta = 0.18 GeV^{2}$. During the
description of $q \bar q$-system we were changing the parameter
$\beta_V$ (when $\beta_V + \beta_S = 0.18 GeV^2)$ to achieve the
agreement with experimental mass splitting of $1^{--}$-states.
Tables 2 - 4 list calculated results. We
also display experimental magnitudes [19]. \\
\begin{table}[2]
\caption{\bf  Hyperfine splitting for the charmonium}
\begin{center}
\begin{tabular}{|c|c|c|c|c|c|c|}
\hline State & $S$-wave & $SD$-waves & [20] & [19] & $E_{SD},$
&$P_D$,\\
& $E_{theor}, MeV$ & $E_{theor}, MeV$ & $E_{theor},
MeV$ & $E_{exp}, MeV$ & \% & \% \\
\hline
$1^1 S_0$ & 2980 & & & 2980 & & \\
\hline
$1^3 S_1$ & 3153 & 3097 & & 3097 & 16 & 0.05  \\
\hline
$1^3 S_1 - 1^1 S_0$ & 173 & 117 & 110 & 117 & & \\
\hline
$2^1 S_0$ & 3642 &  &  & 3590 & & \\
\hline
$2^3 S_1$ & 3759 & 3734 &  & 3685 & 3 & 0.8 \\
\hline
$2^3 S_1 - 2^1 S_0$ & 117 & 92 & 67 & 95 & & \\
\hline
$3^1 S_0$ & 4107 & & & - & & \\
\hline
$3^3 S_1$ & 4208 & 4192 & & 4040 & 1 & 1.3 \\
\hline
$3^3 S_1 - 3^1 S_0$ & 101 & 85 &  & -  &  & \\
\hline
\end{tabular}
\end{center}
\end{table}
\begin{table}[3]
\caption{\bf  Hyperfine splitting for the bottonium}
\begin{center}
\begin{tabular}{|c|c|c|c|c|c|c|}
\hline State & $S$-wave & $SD$-waves & [20] & [19] & $E_{SD},$
&$P_D$,\\
& $E_{theor}, MeV$ & $E_{theor}, MeV$ & $E_{theor},
MeV$ & $E_{exp}, MeV$ & \% & \% \\
\hline
$1^1 S_0$ & 9415 & & & - & & \\
\hline
$1^3 S_1$ & 9462 & 9460 & & 9460 & 2.1 & 0.004  \\
\hline
$1^3 S_1 - 1^1 S_0$ & 47 & 45 & 46 & - & & \\
\hline
$2^1 S_0$ & 9883 &  &  & - & & \\
\hline
$2^3 S_1$ & 9911 & 9911 &  & 10023 & 0.2 & 0.04 \\
\hline
$2^3 S_1 - 2^1 S_0$ & 28 & 28 & 26 & - & & \\
\hline
$3^1 S_0$ & 10201 & & & - & & \\
\hline
$3^3 S_1$ & 10224 & 10224 & & 10355 & 0.1 & 0.1 \\
\hline
$3^3 S_1 - 3^1 S_0$ & 23 & 23 &  & -  &  & \\
\hline
\end{tabular}
\end{center}
\end{table}
\begin{table}[4]
\caption{\bf  Hyperfine splitting for $(u \vec u)$-systems}
\begin{center}
\begin{tabular}{|c|c|c|c|c|c|c|}
\hline State & $S$-wave & $SD$-waves & [20] & [19] & $E_{SD},$
&$P_D$,\\
& $E_{theor}, MeV$ & $E_{theor}, MeV$ & $E_{theor},
MeV$ & $E_{exp}, MeV$ & \% & \% \\
\hline
$1^1 S_0$ & 140 & & & 140 & & \\
\hline
$1^3 S_1$ & 674 & 640 & & 770 & 4 & 0.02  \\
\hline
$1^3 S_1 - 1^1 S_0$ & 534 & 500 & 923 & 630 & & \\
\hline
$2^1 S_0$ & 1134 &  &  & 1300 & & \\
\hline
$2^3 S_1$ & 1564 & 1543 &  & 1450 & 1 & 0.04 \\
\hline
$2^3 S_1 - 2^1 S_0$ & 430 & 409 & 411 & 150 & & \\
\hline
\end{tabular}
\end{center}
\end{table}
\hspace*{0.5cm} Tables 2 - 4 present the mass spectra of
pseudoscalar and vector mesons. They also give the size of
$D$-wave admixture in the wave function of vector meson and
the input of interferential member $E_{SD}$. \\
\hspace*{0.5cm} We also have calculated root-mean-square radii
and results are provided in Table 5.
\begin{table}[5]
\caption{\bf  Squared mean value radius}
\begin{center}
\begin{tabular}{|c|c|c|c|c|}
\hline
$nL$ & [21] & Our results  & [21] & Our results \\
& $\langle r^2_c \rangle^{1/2},$ fm & $\langle r^2_c
\rangle^{1/2}, $fm & $\langle r^2_b \rangle^{1/2},$ fm & $\langle
r^2_b
\rangle^{1/2},$ fm \\
\hline
1 S & 0.43 & 0.433 & 0.24 & 0.256 \\
\hline
2 S & 0.85 & 0.847 & 0.51 & 0.552 \\
\hline
3 S & 1.18 & 1.182 & 0.73 & 0.768 \\
\hline
\end{tabular}
\end{center}
\end{table}
\hspace*{0.5cm} Squared mean value radius 0.7 fm meet the
requirements of quark-antiquark pair creation $(u \bar u)$
(string break). It is necessary modify the potential model taking
into account the opening of a new channel for those conditions
where the given value is overlapped.
\vspace*{0.5cm} \\
{\bf 3. Leptonic decay of heavy quarkonia}
\vspace*{0.5cm} \\
\hspace*{0.5cm} For the leptonic decay widths of two-quark system
we shall consider decay of $^3S$ (vector) states into $e^+ e^-$
pairs. The leptonic decay width of system $Mq \bar q \to e^+ e^-$
is calculated from the  Van Royen Weisskopf  formula [22]
\begin{equation}
\tilde \Gamma (^3S_1 \to e^+ e^-) = \frac{4 \alpha^2_{em}Q^2}
{M_{q \bar q}}|R(0) |^2. \label{eq21}
\end{equation}
Where $M_{q \bar q}$ is mass of vector meson, $Q$ is quark charge,
$\alpha_{em}$ is the fine structure constant and
$R(0)$ is the radial wave function at the origin. \\
\hspace*{0.5cm} The formula (\ref{eq21}) is based on the notion
that constituent quark-antiquark pair annihilates into a single
virtual foton, which subsequently gives rise to a leptonic
pair.\\
\hspace*{0.5cm} The decay widths of the vector $b \bar b$ and  $c
\bar c$ quarkonia into charged lepton [23]
\begin{equation}
\Gamma (^3 S_1 \to e^+ e^-) = \tilde \Gamma (1 - \frac{16
\alpha_s (m_{q}^2 )}{3 \pi}). \label{eq22}
\end{equation}
\hspace*{0.5cm} As Eichten and Quigg have pointed out [24] the
$QCD$ correction reduces the magnitude of $\Gamma$ significantly,
however the amount of reduction is somewhat uncertain . For
vector mesons containing light quarks this
formula leads to paradoxes [25].\\
\hspace*{0.5cm} In paper [2] Motyka and Zalewski have got formula
\begin{equation}
\Gamma_{V \to e^+ e^-} = F(q) \frac{32 \alpha_{s}}{9 M^2_V}
|R(0)|^2, \label{eq23}
\end{equation}
with $F(c) = 4.73 . 10^{-5}$ for charmonium and $F(b) = 2.33 .
10^{-5}$ for bottonium. We have calculated decay widths by Van
Royen-Weisskopf (\ref{eq21}) and by (\ref{eq23}). Table 6 lists
results of calculation.
\begin{table}[6]
\caption{\bf  The leptonic decay widths of heavy mesons}
\begin{center}
\begin{tabular}{|c|c|c|c|c|c|c|}
\hline
State & $SD$-waves & $S$-wave  & [2] & [26] & [27]& [19] \\
& $\Gamma_{theor.}$,keV &  $\Gamma_{theor.}$,keV &
$\Gamma_{theor.}$,keV & $\Gamma_{theor.}$,keV &
$\Gamma_{theor.}$,keV &  $\Gamma_{exp.}$,keV
\\
\hline $J/ \psi 1 S$ & 7.8 (5.41) & 8.2 (5.63) & 4.5 & 4.24 & 8.0
& $5.26 \pm 0.37$  \\
$\psi'2 S$ & 3.7 (2.59) & 4.0(2.79) & 1.9 & 1.81 & 3.7 &
$2.12 \pm 0.18$ \\
$\psi'' 3 S$ & 2.6 (1.82) & 2.9 (2.01) & - - - & 1.22
& - - - & $0.75 \pm 0.15$ \\
\hline $\Upsilon 1 S$ & 1.14 (0.96) & 1.20 (1.01) & 1.36 & 0.85 &
1.7 & $1.32 \pm 0.04$ \\
$\Upsilon' 2 S$ & 0.58 (0.49) & 0.63 (0.53) & 0.59 & 0.38 &
0.8 & $0.52 \pm 0.03$ \\
$\Upsilon'' 3 S$ & 0.44 (0.37) & 0.49 (0.42) & 0.40 & 0.27 &
0.6 & $0.48 \pm 0.08$ \\
\hline
\end{tabular}
\end{center}
\end{table}
The calculations of widths were done with tensor forces and
without tensor forces. Value of the widths which were calculated
by formula (\ref{eq23}) are given in parentheses.
\vspace*{0.5cm} \\
{\bf 4. Conclusion}
\vspace*{0.5cm} \\
\hspace*{0.5cm} The analysis of our results presented  in Table 2
- 4 shows that percentage differences between the theoretical
calculations of mass spectrum for heavy quarkonia and the
experimental results are 1 - 4 \%. Therefore we expect
relativistic correction in the range 1 - 16 \% for charmonium,
and up to 1 \% for bottonium spectrum. Also we have  been able to
describe hyperfine splitting of $c \bar c$- and  $b \bar
b$-quarkonium. For describing mass spectrum of light mesons it is
necessary to use relativistic potential model. We have calculated
hyperfine splitting $u \bar u$ system, too.  \\
\hspace*{0.5cm} The leptonic decay widths suggest that for
charmonium theoretical widths, which were calculated by Van
Royen-Weisskopf formula are systematically higher than
experimental data. But for bottonium we have obtained  lower
values. For $J / \psi$-meson better decay widths are obtained by
means of formula (\ref{eq23}), which takes into account $QCD$
correction. Here the influence of $D$-waves ranges from 4 \% for
ground state up to 25 \% for the second excited state, but for
values calculated by Van Royen-Weisskopf formula it is from 8 \%
up to 50 \%. For $\Upsilon$-meson it is opposite. More exact is
calculation done by (\ref{eq21}), but $QCD$ correction reduced
the results more apparent. Besides, $D$-waves contribute less
than for charmonium: from 4 \% for ground state up to  11 \% for
second excited state. \\
\hspace*{0.5cm} The results obtained show that contribution of
the $D$-waves is impossible to neglect for considered leptonic
decay width of quarkonia.

\end{document}